\documentclass[showpacs,preprint,amsmath,amssymb,aps,11pt]{revtex4}

\usepackage{graphicx}
\usepackage{graphicx,rotating,lscape}

\begin{document}

\def\erf{\mathrm{erf}}

\title{Dynamically generated open charmed baryons beyond the zero range approximation}

\author{C. E. Jim\'{e}nez-Tejero$^1$, A. Ramos$^1$ and I. Vida\~na$^2$}

\affiliation{$^1$Departament d'Estructura i Constituents de la Mat\`eria and
             Institut de Ci\`{e}ncies del Cosmos,
    Universitat de Barcelona, Avda. Diagonal 647, E-08028 Barcelona, Spain}

\affiliation{$^2$Centro de F\'{i}sica Computacional, Department of Physics,
    University of Coimbra, 3004-516 Coimbra, Portugal}

\begin{abstract}
The interaction of the low lying pseudo-scalar mesons with the ground state baryons in the
charm sector is studied within a coupled channel approach using a t-channel vector-exchange
driving force.
The amplitudes describing the scattering of the pseudo-scalar mesons
off the ground-state baryons are obtained by solving the Lippmann--Schwinger equation. We analyze in
detail the effects of going beyond the $t=0$ approximation. Our model
predicts the dynamical generation of several open charmed baryon resonances in different isospin and strangeness
channels, some of which can be clearly identified with recently observed states.
\end{abstract}

\vspace{0.5cm}

\maketitle


\section{Introduction}

A very active topic of research in hadron physics concerns the study and
characterization of resonances, in order to establish whether they qualify
as genuine $q\bar q$ or $qqq$ states or, alternatively, they behave more
as hadron molecules generated dynamically. A series of pioneer works
\cite{ball61,Wyld:1967zz,Dalitz:1967fp,Logan:1967zz,raja72,sie88}, based on a $t$-channel vector meson exchange
force, already predicted a wealth of s-wave baryon resonances generated
by coupled channel dynamics with effective hadronic degrees of freedom rather
than quarks and gluons. The earlier approaches have been readapted in the last
decade to the modern language of chiral lagrangians
\cite{Kai95,Nie01,OR98,Meissner:1999vr,OOR,Oller:2000fj,Nieves:2001wt,Inoue:2001ip,Lutz:2001yb,Garcia-Recio:2002td,ORB02,Ramos:2002xh,Jido:2003cb,Oller:2005ig,Borasoy:2005ie,Borasoy:2006sr,Hyodo:2008xr,Hyodo:2007np}
and many resonances in the light SU(3) sector, which cannot be described properly by quark models \cite{Capstick:1986bm}
unless substantial meson-baryon components are included \cite{Gonzalez:2008en}, have been identified with dynamical
states generated from the interactions of mesons of the pseudoscalar
$0^-$ octet with the $1/2^+$ ground state baryons. Some consequences of these studies, such as
the two-pole nature of the $\Lambda(1405)$,
are confirmed through the analyses \cite{Magas:2005vu,sekihara}
of different experimental reactions \cite{Thomas,Prakhov,Braun:1977wd}.
Note that the basic structure of a molecular-type baryon is quite
different from that implied by the quark models, even when they include the
dressing with meson-baryon components, as in the
$^3P_0$ formalism, where the $qqq$ Hilbert space is coupled to the meson-baryon
Hilbert space through the creation of a $q{\bar q}$ pair
\cite{orsay,Koniuk:1979vy,Tornqvist:1984fy,Tornqvist:1985fi,Blask:1987yv,SilvestreBrac:1991pw}.
In the former case the degrees of freedom are purely hadronic and the
resonances must be seen as pseudo-bound states of two hadrons, while in the
later case the essential component of a baryon is still of three quark nature.
A goal in hadron physics research is to distinguish between both pictures from
thorough analyses of as many properties of the hadron as possible, such as the
mass, width, magnetic moment, form-factors, etc\dots

In recent years it was demonstrated that besides the $s$-wave baryon resonances
many more states can be generated dynamically. Baryon resonances with
$J^P=3/2^-$ were studied based on the leading order chiral lagrangian with
the decuplet $3/2^+$ fields \cite{Kolomeitsev:2003kt,Sarkar:2004jh,Roca:2006sz,Doring:2006pt}.
$D$-wave baryon resonances  were also generated dynamically with vector meson degrees
of freedom in Refs.\ \cite{lutz3,Garcia-Recio:2005hy,Toki:2007ab,vijande,souravbao,osetramos}.
Another promising line of research is the recent interpretation
of low lying $J^P=1/2^+$ resonances as molecular states of two pseudoscalar
mesons and one baryon \cite{alberto,alberto2,kanchan,Jido:2008zz,KanadaEn'yo:2008wm}.
All these results support the so-called hadrogenesis conjecture, formulated a few
years ago by Lutz and Kolomeitsev, according to which
resonances not belonging to the large-$N_c$ ground state of QCD are generated
by coupled-channel dynamics \cite{Lutz:2001yb,lutz1,lutz2,lutz3,lutz4}.

The study of charmed hadrons is receiving an increased attention
thanks to the efforts of a series of collaborations, both at lepton
colliders (CLEO, BELLE, BaBar) and hadron facilities (CDF@Fermilab, PHENIX
and STAR@RHIC, FAIR@GSI). The new results confirm with better
statistics previously seen charmed states but are also giving rise to the
discovery of a large amount of new hadrons
\cite{Aubert:2003fg,Besson:2003cp,Krokovny:2003zq,Choi:2003ue,Acosta:2003zx,2880-Artuso:2000xy,Mizuk:2004yu,Jessop:1998wt,Csorna:2000hw,Chistov:2006zj,Aubert:2007dt,Aubert:2006je,2940-Aubert:2006sp,2880-Abe:2006rz}.
Coupled-channel unitary schemes have been recently extended
to include the charm degree of freedom and have been applied to
the description of open and hidden charm mesons, with the observation that some
states admit a straightforward
interpretation as meson molecules \cite{Kolomeitsev:2003ac,Hofmann:2003je,Guo:2006fu,Guo:2006rp,Gamermann:2006nm,Gamermann:2007fi}.
Similar methods have been used for describing baryons with charm, motivated in
part by a clear parallelism between the behavior of the
$\Lambda(1405)$ in the $C=0, S=-1$ sector with the $\Lambda_c(2595)$
in the $C=1$ and $S=0$ one
\cite{Tolos:2004yg,Lutz:2003jw,lutz5,Hofmann:2005sw,Hofmann:2006qx,Mizutani:2006vq}.
To be consistent with the spin-flavour Heavy Quark Symmetry that develops in
this heavy sector \cite{IW89,Ne94,MW00},
the vector mesons and $J=3/2^+$ baryons
have recently been included in the basis of meson-baryon states, employing
a static spin-flavour SU(8) scheme \cite{GarciaRecio:2008dp}
similar to that developed in the light sector
\cite{Garcia-Recio:2005hy,Toki:2007ab}.
Treating the $D$ and $D^*$ mesons on an equal footing basis has lead
to the observation that some of the dynamical generated states,
such as the $\Lambda_c(2595)$, have mostly a $D^* N$ composition rather than the $DN $ molecular
nature. In any case, the observation that some of the
dynamically generated charmed hadrons can be readily identified with observed resonances, such as the $J^P=1/2^-$ $\Lambda_c(2595)$ or the
$J^P=3/2^-$ $\Lambda_c(2625)$ charmed baryons, sustains the hadrogenesis conjecture \cite{Lutz:2001yb,lutz1,lutz2,lutz3,lutz4}.

Apart from the different basis of states included in the models,
a common feature of all the previous works is the use of an interaction based on the
$t$-channel exchange of vector mesons, as
the driving force for the $s-$wave scattering of pseudo-scalar
mesons off ground state baryons. The limit $t \to 0$ is applied,
leading to a vector type WT zero-range interaction.
This procedure is justified for on-shell meson-baryon transitions, $MB
\to M^\prime
B^\prime$, which are diagonal ($M^\prime B^\prime = MB$) and hence the
value of $t$ is small as long as one is not too far from threshold.
It also holds for non-diagonal amplitudes  ($M^\prime B^\prime \neq MB$)
that
show a moderate difference of masses
between the initial and final mesons and baryons involved,
as is the case of meson-baryon
scattering within the light SU(3) world. However, the coupled channel dynamics
in the heavy sector finds also charm-exchange processes for which the difference of
masses between the external mesons are comparable with the mass of the charmed
vector meson being exchanged. This clearly signals the breakdown of the
zero-range approximation which is no longer reliable for these non-diagonal
transitions. While one may still argue that many
of the dynamically generated states are triggered by a single
dominant meson-baryon interaction component and, hence, their energy can be
well estimated by the pole position of an uncoupled calculation involving only
diagonal amplitudes, the corresponding width will however be determined
by non-diagonal amplitudes
and will therefore depend on whether the $t=0$ approximation
is implemented or not.
Moreover, it is well known that some resonances owe their origin to
a particularly strong coupling between different channels, hence involving
non-diagonal transitions, in which case
the $t=0$ approximation is not at all appropriate for these states.

In the present work we study the charmed baryon resonances obtained
dynamically from the interaction of the low lying pseudo-scalar mesons
with the ground state baryons within a coupled channel approach, using
the full $t$-dependence of the t-channel vector-exchange driving term,
instead
of the $t=0$
approximation.
We
incorporate the $t$-dependence within a general four-dimensional
integration scheme, which we reduce to a three-dimensional equation of the
``Lippmann-Schwinger" type, and
analyze in detail the effects of going beyond
the $t=0$ approximation within this scheme.

The paper is organized as follows.
In Sec.\ \ref{sec:Formalism} we present
the formalism, showing the details of the kernel employed and the
equation used to obtain the scattering amplitudes.
Our results for the properties of the baryon resonances with charm
in various strangeness and isospin cases are shown in Sec.\ \ref{sec:Results},
where we also compare with the $t=0$ results.
 A summary of our conclusions is
presented in Sec.\ \ref{sec:Conclusions}.


\section{Formalism}
\label{sec:Formalism}

Following the original work of Hofmann and Lutz \cite{Hofmann:2005sw}, we
identify a t-channel exchange of vector mesons as the driving force for the s-wave scattering. In their original model,
Hofmann and Lutz exploited the universal vector meson coupling hypothesis. They considered the t-channel exchange of vector mesons between pseudoscalar
mesons in $16$-plet and baryons in $20$-plet representations in such a way to respect chiral symmetry for the
light meson sector. The Weinberg--Tomozawa type interaction is recovered in the zero range limit, ({\it i.e.,} $t \to 0$ and see Eqs.\ (6) and (8) of Ref.\ \cite{Hofmann:2005sw}
for details). The scattering kernel has the form
\begin{equation}
V^{(I,S,C)}_{ij}(k_i,q_i,k_j,q_j)=\frac{g^2}{4}\sum_{V \in [16]} C^{(I,S,C)}_{ij;V}\bar u(p_j)\gamma^\mu
\left(g_{\mu\nu}-\frac{(q_i-q_j)_\mu (q_i-q_j)_\nu}{m^2_V}\right)
\frac{1}{t-m^2_V}(q_i+q_j)^\nu u(p_i) \ ,
\label{eq:sk}
\end{equation}
where the sum runs over all vector mesons of the SU(4) $16$-plet, $(\rho$, $K^*$, $\bar K^*$, $\omega$, $\phi$,
$D^*$, $D_s^*$, $\bar D^*$, $\bar D_s^*$, $J/\Psi)$, $m_V$ is the mass of the exchanged vector meson, $g$ is the universal vector meson coupling constant, $k_i, q_i,
k_j$ and $q_j$ are the four momenta of the incoming and outgoing baryon and meson, and the coefficients
$C^{(I,S,C)}_{ij;V}$ denote the strength of the interaction in
the different sectors $(I,S,C=1)$ and channels $(i,j)$. The value of $g=6.6$ reproduces the decay width of the rho meson \cite{pdg}.

Assuming that $|t/m^2_V| << 1$, the $t$-dependence of the scattering kernel can
be neglected, giving rise to the zero-range approximation. The s-wave projection of the scattering kernel under such
approximation is easily obtained, and in the center--of--mass frame it takes the form
\begin{eqnarray}
V_{ij,l=0}^{(I,S,C)}(k_i,q_i,k_j,q_j) = - N \frac{g^2}{4}
\sum_{V \in[16]}
&&\frac{C^{(I,S,C)}_{ij;V}}{m^2_V}
\left(\omega(|\vec k_i|)+E(|\vec k_i|)+\omega(|\vec k_j|)+E(|\vec k_j|)-M_i-M_j \right. \nonumber \\
 &&\phantom{\frac{C^{(I,S,C)}_{ij;V}}{m^2_V}}~~ - \left. \frac{m^2_j-m^2_i}{m^2_V}(M_i-M_j)\right) \ ,
\label{eq:t0}
\end{eqnarray}
where $m_i,m_j,M_i,M_j$ are the masses of the incoming and outgoing mesons and
baryons, and $\omega_i(|\vec k_i|),\omega_j(|\vec k_j|),
E_i(|\vec k_i|), E_j(|\vec k_j|)$ their corresponding energies, which
have been taken to be their on-shell values.
The factor $N=[(E(|\vec k_i|)+M_i)(E(|\vec k_j|)+M_j)/(4M_i M_j)]^{1/2}$
comes from the normalization of the Dirac spinors. Note that for
physical (fully on-shell) transition amplitudes
one has $\omega_i(|\vec k_i|)+E_i(|\vec k_i|)=
\omega_j(|\vec k_j|)+E_j(|\vec k_j|)=\sqrt{s}$, in which case one
recovers the familiar expression of the commonly used Weinberg--Tomozawa
interaction. The last term in Eq.~(\ref{eq:t0})
is usually ignored in most works, to be more
consistent with the $t=0$ approximation applied in the denominator of the
meson-exchange propagator, as noted in Ref.~\cite{Mizutani:2006vq}.
In any case, its consideration introduces only
minor corrections \cite{Hofmann:2005sw}.

\begin{figure}[ht!]
\includegraphics[height=8.5cm,angle=0]{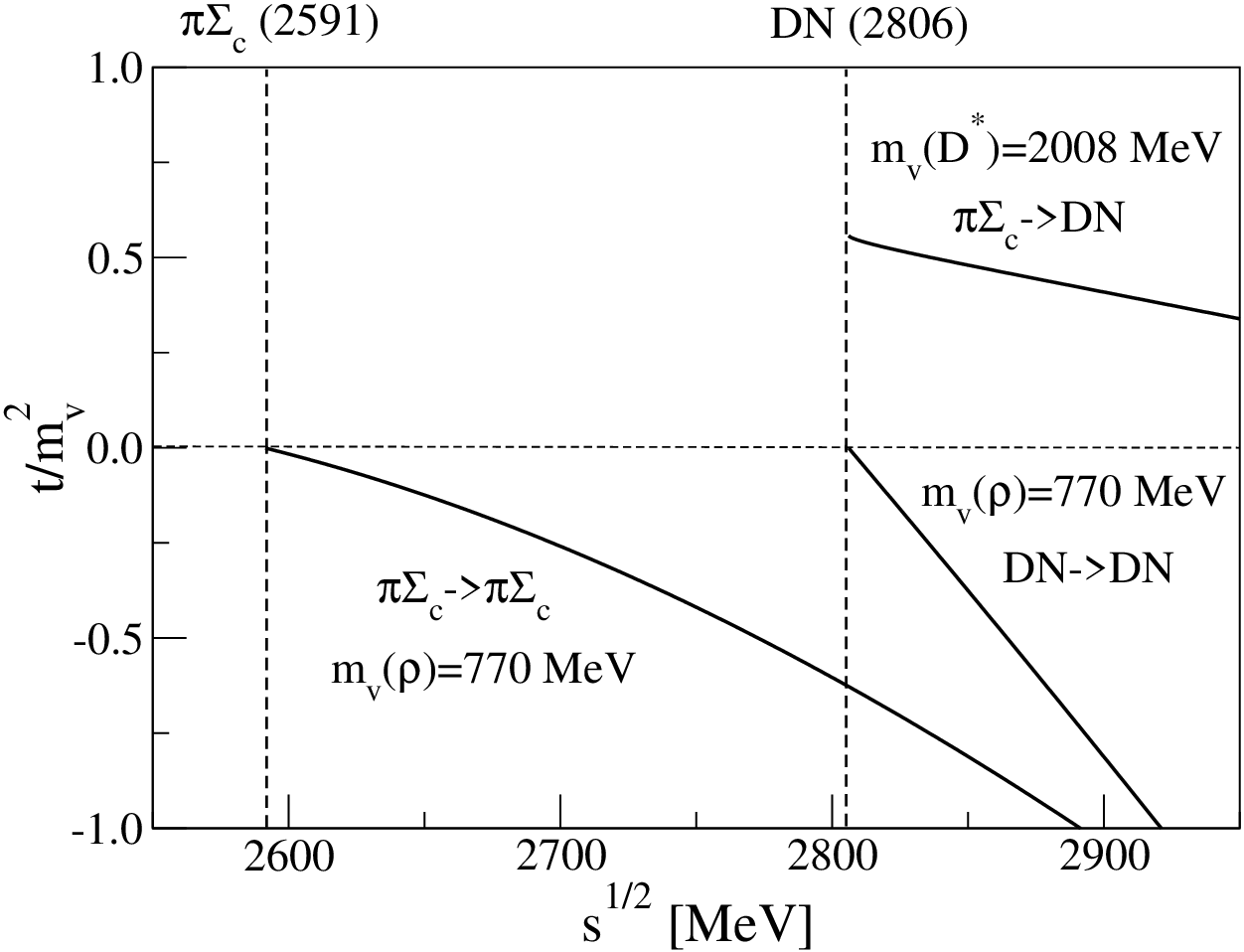}
\vspace{0.25cm}
\caption{Dependence on the
center-of-mass energy $\sqrt{s}$ of the four momentum tranfer $t/m_V^2$ for
$\cos\theta=-1$ and for different transition amplitudes.}
\label{fig:tm2}
\end{figure}

In order to illustrate the validity of this approach we show
in Fig.~\ref{fig:tm2} the value of $ t/m^2_V$ for $\cos\theta=-1$
as a function of $\sqrt{s}$,
where $m_V$ is the mass of a representative
meson exchanged, which we take to be the $\rho$ meson mass for diagonal transitions
and
the $D^*$ meson mass for charm exchange ones.
The range of energies goes roughly between the
$\pi\Sigma_c$ and $DN$ thresholds, thereby
covering the region of the $J^P=1/2^-$ resonance
$\Lambda_c(2595)$, which is a prime example of a dynamically
generated open charmed baryon state in various
approaches \cite{lutz5,Tolos:2004yg,Lutz:2003jw,Hofmann:2005sw,Hofmann:2006qx,Mizutani:2006vq}.
It also expands beyond the $DN$
threshold for about 300 MeV in order to explore the energy region
that will be relevant
in future studies of the $D$-meson self-energy in a nuclear medium.
As one can see, the value of $t/m^2_V$ is only close to zero for diagonal transitions around their corresponding energy threshold,
and its size can be comparable to one at the energies of interest. For the non-diagonal
$\pi\Sigma_c \to DN$ transition, $t/m^2_V$ never goes to zero and acquires values of the order of 0.5.

\newpage

The results of Fig.~\ref{fig:tm2} point clearly to the need for exploring the effects of going beyond
the $t=0$ approximation, an attempt that is taken in the present work by
considering
the full $t$-dependence of the scattering kernel given by Eq.\ (\ref{eq:sk}).
Performing the
s-wave projection, we obtain the following analytic expression:
\begin{equation}
V_{ij,l=0}^{(I,S,C)}(\vec k_i,\vec k_j)=
N\frac{g^2}{8}
\sum_{V \in[16]}
C^{(I,S,C)}_{ij;V} \left[ \frac{2\beta}{b}+\frac{\alpha b-\beta a}{b^{2}}\ln\left(\frac{a+b}{a-b}\right)\right] \ ,
\label{eq:t}
\end{equation}
with $a, b, \alpha$ and $\beta$ being
\begin{eqnarray}
a &=& m^2_i+m^2_j-2\omega_i(|\vec k_i|) \omega_j(|\vec k_j|)-m^2_V  \, \nonumber \\
b &=& 2 |\vec k_i| |\vec k_j| \, \nonumber \\
\alpha &=& \Omega_i(|\vec k_i|)+\Omega_j(|\vec k_j|)
-M_i-M_j-\frac{m^2_j-m^2_i}{m^2_V}(\Omega_j(|\vec k_j|)-\Omega_i(|\vec k_i|)+M_i-M_j) \, \nonumber \\
\beta &=& \frac{|\vec k_i||\vec k_j| }{(E_i(|\vec k_i|)+M_i)(E_j(|\vec k_j|)+M_j)}
\left(\Omega_i(|\vec k_i|)+\Omega_j(|\vec k_j|)+M_i+M_j \phantom{-\frac{m^2_j-m^2_i}{m^2_V}}
\right. \nonumber \\
&& \phantom{\frac{|\vec k_i||\vec k_j|}{(E(|\vec k_i|)+M_i)(E(|\vec k_j|)+M_j)} }
\left. -\frac{m^2_j-m^2_i}{m^2_V}(\Omega_j(|\vec k_j|)-\Omega_i(|\vec k_i|)+M_j-M_i)\right)
\ ,
\end{eqnarray}
where we have defined $\Omega(|\vec k|)\equiv\omega(|\vec k|)+ E(|\vec k|)$.

The $t=0$ expression of the s-wave interaction is recovered by expanding the
logarithm of Eq.~(\ref{eq:t}) in the limit $b/a\to 0$ up to the linear term in
$b/a$, and setting $a = -m_V^2$.
As one can infer from the values of $t/m_V^2$ displayed in
Fig.~\ref{fig:tm2}, keeping the $t$-dependence in the denominator
of the kernel [see Eq.~(\ref{eq:sk})] will decrease the
strength of the diagonal transitions, since in this case $t<0$. Therefore, in
order to reproduce a given resonance found in local models, the present
approach will in general need to compensate the lack of strength with a higher
value of the cut-off momentum used to regularize the loop integrals. On the
other hand, non-diagonal amplitudes, responsible mainly for the decay width of
the dynamically generated states, will be enhanced since they are characterized
by a positive time-like $t$ value due to the large mass difference between the
mesons and baryons involved in the transition. As a consequence, our resonances
will be wider than those found in the local models.

Retaining the $t$-dependence in the kernel implies additional analytical
structures \cite{Lahiff:2002wp}
 which prevent us from obtaining the scattering
amplitudes by
solving the Bethe-Salpeter (BS) equation using on-shell amplitudes only.
Instead, we
incorporate the $t$-dependence within a more general four-dimensional
integration scheme, which we reduce to a three-dimensional equation of the
``Lippmann-Schwinger" type. To this end, for a given value of the
total scattering energy,
we will evaluate the transition potential between any arbitrary pair of relative
momenta within the cut-off value,
keeping the dependence on the momentum transfer in the corresponding matrix
element. This procedure follows the same spirit of
 the usual meson-exchange models of the
$NN$ interaction \cite{machleidt}, also applied to
meson-baryon scattering models \cite{muellerKN,haidenbauerKN}.
We note that
keeping the full $t$ dependence in the exchanged meson
propagator also implies that retardation effects are implemented as done
in the three-dimensional reduction of the Bethe-Salpeter equation of
Ref.~\cite{gross}, which is different than the
prescription based on time-dependent perturbation theory \cite{machleidt}.
Both choices differ in the way off-shell effects are implemented and their
differences in the $NN$ sector show up especially in the
contributions of the two-meson exchange box diagrams \cite{machleidt} not included
in the present model. In any case, observables can
be matched to experimental data with either choice of retardation effects
by selecting appropriate values of the renormalization
parameters.

We note that Eqs.~(\ref{eq:sk}) to (\ref{eq:t}) assume infinitely lived
(zero width) exchanged vector mesons, while some of them have large widths
 due to their strong decay into a pair of mesons, like the $\rho$ meson
contributing to diagonal channels. We have checked, however, that the
value of $t$ is never larger than the square of the minimum
energy required for the meson to decay, namely $(2m_\pi)^2$ in the case of
$\rho$ exchange or $(m_\pi + m_K)^2$ for $K^*$. In other words,
the mesons being exchanged in this problem are largely off-shell and they
will be treated as stable particles.

Once the scattering kernel has been constructed, we can obtain the $T$-matrices
which describe the scattering of the pseudo-scalar meson fields off the baryon
fields by solving the well known Lippmann--Schwinger equation
\begin{equation}
T^{(I,S,C)}_{ij,l=0}(\vec k_i,\vec k_j,\sqrt{s})=V^{(I,S,C)}_{ij,l=0}(\vec k_i,\vec k_j)
+\sum_{m}\int\frac{d\vec k}{(2\pi)^3}F(|\vec k|)V^{(I,S,C)}_{im,l=0}(\vec k_i,\vec k)
J_m(\sqrt{s},\vec k)T^{(I,S,C)}_{mj,l=0}(\vec k,\vec k_j,\sqrt{s}) \,
\end{equation}
where
\begin{equation}
J^{(I,S,C)}_m(\sqrt{s},\vec k)=\frac{M_m}{2E_{m}(|\vec k|)\omega_{m}(|\vec k|)}
\frac{1}{\sqrt{s}-E_{m}(|\vec k|)-\omega_{m}(|\vec k|)+i\eta} \ .
\label{eq:prop}
\end{equation}
We have introduced a dipole-type form factor $F(|\vec k|)$
\begin{equation}
F(|\vec k|)=\left(\frac{\Lambda^2}{\Lambda^2+|\vec k|^2}\right)^2
\end{equation}
to regularize the integral. This form is typically adopted in studies of
hadron-hadron interactions within the scheme of Lippmann-Schwinger-type
equations in the light flavour sector \cite{machleidt}.
The value of the cut-off $\Lambda$ is a free parameter of our model.
Given the limited amount of data for charmed baryon resonances, and in order to
simplify the analysis,
the cut-off  $\Lambda$ is adjusted to the position of a
well known
$J^P=1/2^-$ state in a particular isospin and strangeness sector,
 and
the same value is used for the other sectors explored in this work.
We will also investigate the effect of a gaussian-type form factor, as well
as the dependence of our results on the value of the cut-off employed.

Note that the approaches based on the Bethe--Salpeter
equation solved with on-shell amplitudes ignore
the off-shellness (momentum dependence)
of the kernel and scattering amplitude in the loop function.
Actually, the on-shell factorization can only be justified within the $WT$ form
of the potential which is obtained after
applying the $t \to 0$ limit.

Another aspect worth commenting is the different extrapolation of the kernel at
subthreshold energies. While the kernel of the on-shell BS approaches
depends on $\sqrt{s}$ in the form given by Eq.~(\ref{eq:t0})
upon replacing $\omega_i(|\vec k_i|)+E_i(|\vec k_i|)+\omega_j(|\vec k_j|)+E_j(|\vec k_j|)$
by $2 \sqrt{s}$, the potential used in the
tri-dimensional Lippmann--Schwinger equation depends essentially on the incoming
and outgoing tri-momenta. These are always taken real quantities in our approach, hence the sum
of the four
single particle energies is always larger than the sum of the two meson-baryon
thresholds involved in the transition. An explicit
dependence on $\sqrt{s}$ is only implemented in the meson-baryon intermediate
propagator of the Lippmann--Schwinger equation that determines the scattering amplitudes.
Therefore, at subthreshold energies, the factor in the numerator of the kernel
used in on-shell BS approaches is smaller than in the present work. This in
part compensates the enhancement in diagonal transitions
associated to the $t \to 0$ limit. In any case,
the free parameters of the model (cut-off values)
can finally be conveniently fine-tuned to adjust the energy position of a
well known  resonance, as is usually done for the case of
the $\Lambda_c(2595)$ appearing about 200 MeV below the threshold of the channel $DN$
to which it couples very strongly. In this respect, the differences between
the present work and the on-shell BS approaches will be made more evident
in properties tied to non-diagonal transitions which are reduced in
the $t\to0$ limit, as is the case of the resonance widths.

In order to associate a given enhancement of the scattering amplitude to a
resonance, we look for a characteristic pole in the unphysical sheet of the
complex energy plane. Our prescription of unphysical sheet is such that,
whenever the real part of the complex energy crosses a meson-baryon threshold
cut, the sign of the on-shell momentum is changed for this channel and for
the already opened
ones, as described in detail in Ref.~\cite{Logan:1967zz}.
Once a pole $z_R$ is found, its value determines
the Breit-Wigner mass ($M={\rm Re\,}z_R$) and width ($\Gamma=2\,{\rm
Im\,}z_R$) of the resonance, as seen from real energies,
if the pole is not too far from the real axis. The
couplings of the resonances into the meson-baryon components of a given sector
are obtained from the residues of the scattering amplitude since, close to the
pole, it can be parameterized in the form:
\begin{equation}
T_{ij,l=0}^{I,S,C}(\vec k_i,\vec k_j,z)=\frac{g_ig_j}{z-z_R} \ .
\label{eq:pole}
\end{equation}
Note that, as it stands, the value of the coupling constants of
Eq.~(\ref{eq:pole}) depend on the particular momentum values chosen  in the
evaluation of the $T$-matrix element. Since we are only interested on the
size of the couplings relative to the various channels, we take the prescription
of evaluating them for the case
$\vec k_i=\vec k_j=0$.


\section{Results and Discussion}
\label{sec:Results}

One of the main interests of this paper is to study the effects of going beyond
the $t=0$ approximation in the kernel, as has been customarily done. We have
already anticipated in the previous section that, in the case of coupled
channel problems involving light and heavy flavored particles, this
approximation is not always justified.

\begin{table}[htbp]
    \setlength{\tabcolsep}{0.1cm}
\begin{center}
\begin{tabular}{c|cccccccccccccccc}
\hline
$(\frac{1}{2},-3,1)$ & &&& & & & & $\bar{K} \Omega_c$ &&&&& & & \\

\hline
 $(0,-2,1)$ & & & & &$\bar{K}\Xi_c$ &$\bar{K}\Xi'_c$  &$D\Xi$
 &$\eta\Omega_c$ &$\eta'\Omega_c$ &$\bar{D}_s\Omega_{cc}$ &$\eta_c\Omega_c$ & & & & & \\

\hline
 $(1,-2,1)$ & & &&& & &$\pi\Omega_c$ &$K\Xi_c$  &$K\Xi'_c$&$D\Xi_c$ & & & & & & \\

\hline
 $(\frac{1}{2},-1,1)$ & $\pi \Xi_{c}$  & $\pi \Xi'_{c}$  & $\bar{K}\Lambda_{c}$  & $ \bar{K}\Sigma_{c}$
 & $D\Lambda $ & $\eta\Xi_{c} $ & $D\Sigma $ & $\eta\Xi'_{c}$ & $K\Omega_{c} $
 & $ D_{s}\Xi$ & $\eta'\Xi_{c}$ &$\eta'\Xi'_{c}$ &$\eta_{c}\Xi_{c}$ & $\bar{D}_{s}\Xi_{cc} $
 & $\bar{D}\Omega_{cc} $ &$\eta_{c}\Xi'_{c}$    \\

\hline
 $(\frac{3}{2},-1,1)$ & & &&&& &$\pi \Xi_{c}$  & $\pi \Xi'_{c}$  & $\bar{K}\Sigma_{c}$ & $D\Sigma$ & & & && &\\

\hline
 $(0,0,1)$   & & & & $\pi \Sigma_c$ & $D N$ & $\eta \Lambda_c$ & $K\Xi_c$  & $K\Xi_c'$ & $D_s\Lambda$ & $\eta'\Lambda_c$ &
$\eta_c\Lambda_c$ & $\bar D\Xi_{cc}$ & & & \\

\hline
$(1,0,1)$   & & & & $\pi \Lambda_c$ & $\pi \Sigma_c$ & $D N$ & $K\Xi_c$ & $\eta\Sigma_c$ & $K\Xi_c'$ & $D_s\Sigma$ &
$\eta'\Sigma_c$ & $\bar D\Xi_{cc}$ & $\eta_c\Sigma_c$ & & \\

\hline
$(2,0,1)$ & &&& & & & & $\pi \Sigma_c$ &&&&& & & \\

\hline
$(\frac{1}{2},1,1)$   & &&& & & &$K \Lambda_c$ &$D_{s}N$ & $K \Sigma_c$ &&&& & & \\

\hline
$(\frac{3}{2},1,1)$   & &&& & & & & $K \Sigma_c$ &&&&& & & \\

\hline
\end{tabular}
\end{center}
\caption{Coupled-channel meson-baryon states with charm $C=1$ and
all possible combinations of
isospin, strangeness (I,S).}
\label{tab:tab1}
\end{table}

All the possible sectors with charm $C=1$ that can
be built from the s-wave scattering of pseudoscalar mesons with $J^P=1/2^+$
baryons are shown in Table~\ref{tab:tab1}, together with the corresponding
meson-baryon coupled channels. In
this work, we will first study the cases in which some resonance with either
$J^P=1/2^-$ or unknown spin-parity has already been observed.
This includes the sectors with isospin, strangeness
quantum numbers $(I,S)= (0,0)$, $(1,0)$ and $(1/2,-1)$, corresponding
respectively to $\Lambda_c$, $\Sigma_c$ and  $\Xi_c$ states, the experimental
information of which is gathered in Table~\ref{tab:exp}. We will next explore
the sector $(I,S)=(0,-2)$ of the $\Omega_c$ states which so far has no
experimental evidences for $J^P=1/2^-$ states. Finally, we will comment on the
$(I,S)=(1/2,1)$ sector that can only be realized with the presence of 5 quarks.

\begin{table}[htbp]
\begin{center}
\begin{tabular}{|c|c|c|c|c|c|}
\hline Resonance [MeV] & $I(J^P)$ & Width [MeV] & Decay modes & Status \\
\hline $\Lambda_c(2595)^+$&$0(\frac{1}{2}^-)$&$3.6^{+2.0}_{1.3}$&$\Lambda_c\pi\pi$, $\Sigma_c\pi$ & *** \\
\hline $\Lambda_c(2765)^+$ or $\Sigma_c(2765)$&$?(?^?)$&$\sim 50$&$\Lambda_c\pi\pi$&* \\
\hline $\Lambda_c(2940)^+$& $0(?^?) $&$17^{+8}_{-6}$& $ND$, $\Sigma_c \pi$& *** \\
\hline $\Sigma_c(2800)$&$1(?^?)$&$75^{+22}_{-17}(\Sigma_c^{++})$,
$62^{+60}_{-40}(\Sigma_c^+)$, $61^{+28}_{-18}(\Sigma_c^0)$ &$\Lambda_c\pi$&*** \\
\hline $\Xi_c(2790)$&$\frac{1}{2}(\frac{1}{2}^-)$&$<15 (\Xi_c^+)$, $<12 (\Xi_c^0)$
&$\Xi'_c\pi$&***\\
\hline $\Xi_c(2930)$&$?(?^?)$&$36\pm 13$&$\Lambda_c K$&*\\
\hline $\Xi_c(2980)$&$\frac{1}{2}(?^?)$&$26\pm 7(\Xi_c^+)$, $20\pm 7(\Xi_c^0)$&
$\Lambda_c \bar{K}\pi, \Sigma_c \bar{K}$&*** \\
\hline $\Xi_c(3055)$&$?(?^?)$&$17\pm 13$&
$\Lambda_c \bar{K}\pi, \Sigma_c \bar{K}$&** \\
\hline $\Xi_c(3080)$&$\frac{1}{2}(?^?)$&$5.8 \pm 1.0 (\Xi_c^+)$, $5.6\pm 2.2
(\Xi_c^0)$&
$\Lambda_c \bar{K}\pi, \Sigma_c \bar{K}, \Sigma_c^* \bar{K}$&*** \\
\hline $\Xi_c(3123)$&$?(?^?)$&$4\pm 4$&$\Sigma_c^* \bar{K}$&* \\
\hline
\end{tabular}
\end{center}
\caption{Masses, widths, decay modes and status of experimental charmed
baryon resonances with $J^P=1/2^-$ or unknown.}
\label{tab:exp}
\end{table}

\subsection{$\Lambda_c$ resonances (I=0,S=0,C=1) sector}

\begin{figure}[ht!]
   \includegraphics[height=7cm,angle=0]{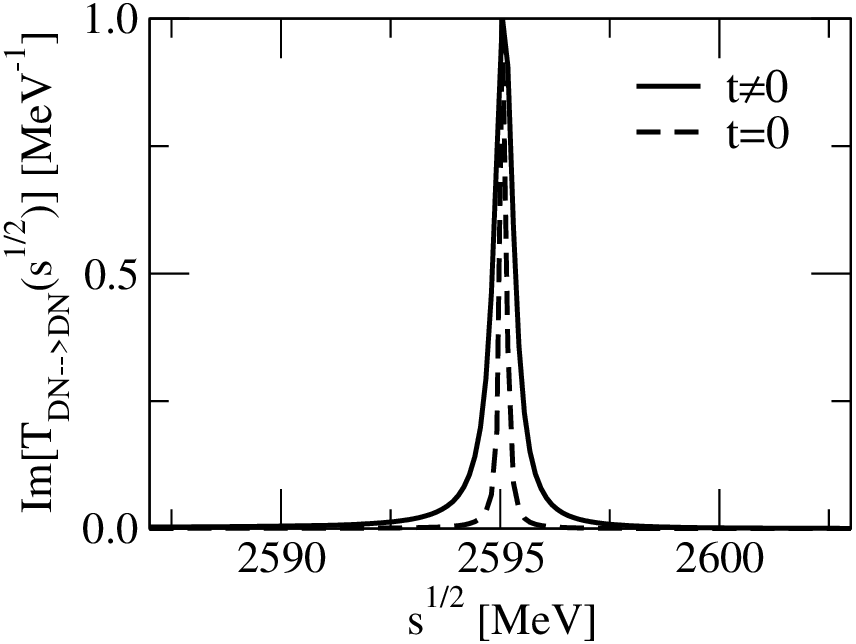}
   \vspace{0.15cm}
   \caption{
   Imaginary part of the scattering amplitude of the elastic process
  $DN \rightarrow DN$ in the $(I,S,C)=(0,0,1)$ sector as a function of
  $\sqrt{s}$ for the finite range interaction (solid line) and the
  zero range approximation (dashed line). The incoming and outgoing
  relative momenta $\vec{k}_i$ and $\vec{k}_j$ have been taken equal to
  $0$.}
   \label{fig:fig1}
\end{figure}

In this sector
there exists a  three-star narrow
$J=1/2^-$ resonance, the $\Lambda_c(2595)$, which has been extensively studied in
various works \cite{Tolos:2004yg,Lutz:2003jw,Hofmann:2005sw,Hofmann:2006qx,Mizutani:2006vq}.
We start by comparing in Fig.~\ref{fig:fig1} the
results obtained using our non-local kernel with those taking the limit
$t\to 0$. We represent the
imaginary part of the scattering amplitude of the elastic process
$DN \rightarrow DN$, as a function of
$\sqrt{s}$ for zero incoming and outgoing relative momentum values.
We can
see that, by adjusting the cut-off value conveniently,  both models of the
kernel can generate this state dynamically. However,
the zero range approximation needs a cut-off value of $\Lambda=553$ MeV
while the
finite range interaction requires a substantially larger value of
$\Lambda=903$~MeV.

This is easily understood from the fact that the $DN \to DN$ diagonal matrix
elements of the non-local potential, largely responsible for generating the
resonance, are smaller in magnitude than those of the local one. The large
difference between the cut-off momentum values is just a reflection of the
importance of the non-local terms in this problem. Once the $\Lambda_c(2595)$ resonance is
conveniently located to its experimental position by both prescriptions, there
remain substantial differences in its width. The local potential produces a
very narrow resonance, of width 0.15 MeV, while the resonance generated by the
finite range potential has a width of 0.5 MeV, closer to the empirical value of
$3.6 + 2.0 -1.3$ MeV. Again, this is due to the different magnitude of the
non-diagonal matrix elements $DN\to \pi \Sigma_c$, which are larger in the
finite range approach. We note that our model does not consider the three-body
decay channel $\Lambda_c \pi\pi$ which  already represents almost one third of
the decay events \cite{pdg}. We also observe that the results obtained with
the low cut-off
in the approximation $t=0$ for the $\Lambda_c(2595)$ agree (mass,
width and couplings) with former studies of meson-baryon resonances of
Hofmann and Lutz in Ref.\ \cite{Hofmann:2005sw} and Garcia-Recio {\it et al.,}
in Ref.\ \cite{GarciaRecio:2008dp}.

Our search of resonances in this sector produces two states that are listed in
Table~\ref{tab:tb6}, together with their widths and couplings to the various
meson baryon states. We immediately see that the $\Lambda_c(2595)$ is basically
a $DN$ state which couples very weakly to its only
possible decaying channel $\pi \Sigma_c$, thereby explaining its narrowness.
We obtain an even narrower
resonance at 2805 MeV, which is a $K\Xi_c$ bound system, a state also found
around the same energy in Ref.\ \cite{Hofmann:2005sw} and  Ref.\ \cite{GarciaRecio:2008dp}.
Note that this resonance
couples non-negligibly to $DN$ and, was its location be moved upward in energy by
20--30 MeV with a slight change of the cut-off parameter, it could explain
part of the structures seen below 2.85 GeV in the $D^0 p$ invariant mass
spectrum measured by the BaBar collaboration \cite{2940-Aubert:2006sp}.
Table~\ref{tab:tb6} also shows the results obtained with the local $t=0$  model.
We remark that, in spite of the fact that the second resonance appears
at a higher energy, 2827 MeV, its width is narrower than in the finite-range model,
confirming the trend observed for the $\Lambda_c(2595)$.

\begin{table}[ht!]
\begin{center}
\begin{tabular}{|c|c|c||c|c|}
\hline
\multicolumn{5}{|c|}{$(I=0,S=0,C=1)$}\\
\hline
$\Lambda$ [MeV]&\multicolumn{2}{|c||}{ $903$ $(t\ne0)$}&
\multicolumn{2}{|c|}{ $553$ $(t=0)$} \\
\hline $M$ [MeV] & $2595$ & $2805$& $2595$ & $2827$ \\
\hline $\Gamma$ [MeV] &$0.5$&$0.01$ &$0.15$&$0.006$\\
\hline
\multicolumn{5}{|c|}{Couplings $|g_i|$}\\
\hline ${\pi\Sigma_{c}}(2591)$ &$0.44$&$0.001$&$0.21$&$0.002$\\
\hline ${DN}(2806)$ & $16.34$&$0.23$&$18$&$0.03$\\
\hline ${\eta\Lambda_{c}}(2832)$&$0.58$&$1.92$&$0.22$&$1.60$\\
\hline ${K\Xi_{c}}(2963)$ & $0.71$ & $3.76$& $0.23$&$3.22$\\
\hline ${K\Xi'_{c}}(3070)$ &$0.32$ &$0.004$& $0.01$ &$0.13$\\
\hline ${D_{s}\Lambda}(3085)$ & $8.16$ & $ 0.18$&$8.54$&$0.07$\\
\hline ${\eta'\Lambda_{c}}(3243)$ & $0.96$ & $0.01$ &$0.54$&$0.003$\\
\hline ${\eta_{c}\Lambda_{c}}(5265)$ &$2.83$&$0.02$& $1.74$ & $0.007$\\
\hline ${\bar{D}\Xi_{cc}}(5307)$ &$0.07$&$0.96$& $0.03$ & $0.48$\\
\hline
\end{tabular}
\end{center}
\caption{Masses, widths and couplings of the resonances in the
$(I,S,C)=(0,0,1)$ sector, for the non-local $(t\ne0)$ and local
$(t = 0)$
models.}
\label{tab:tb6}
\end{table}

\begin{figure}[ht!]
   \includegraphics[height=6cm,angle=0]{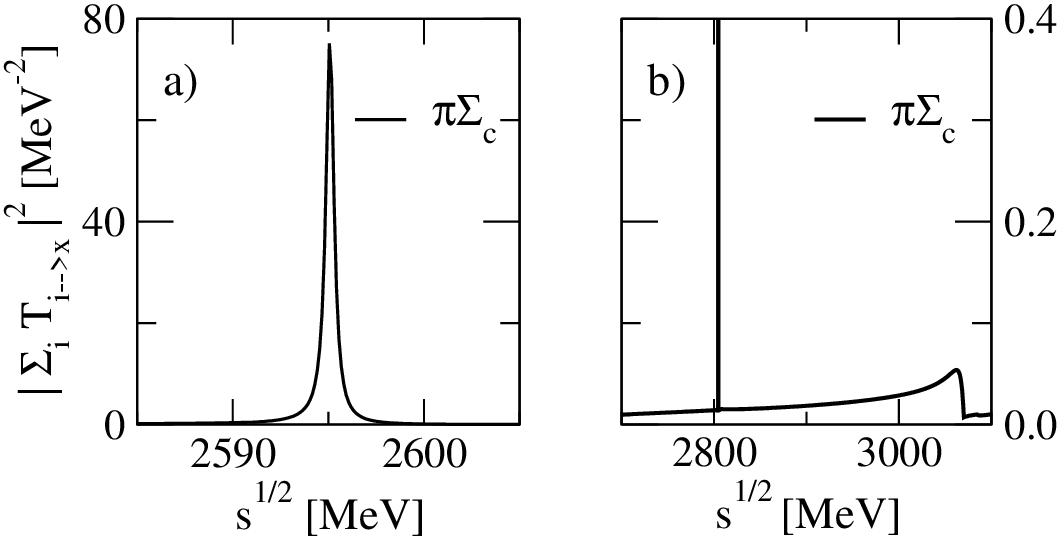}
   \vspace{0.25cm}
   \caption{Modulus square of the coherent sum of all transition amplitudes going to
  any of the possible final meson-baryon decaying channels, as a function of
  $\sqrt{s}$, for the $(I,S,C)=(0,0,1)$ sector. Our results are split into
  two panels, covering different energy regions and having different
  energy scales, to better visualize the properties of each state. The
  incoming and outgoing relative momenta $\vec{k}_i$ and $\vec{k}_j$ have
  been taken equal to $0$.}
   \label{fig:fig2}
\end{figure}

We represent in Fig.~\ref{fig:fig2} the modulus square of the
coherent sum of all transition amplitudes
going to a final
meson-baryon state to which the resonances can decay, namely:
\begin{equation}
\sum_{M^\prime B^\prime} \mid C^R_{M^\prime B^\prime}
T_{M^\prime B^\prime \to M B}(\sqrt{s}) \mid^2 \ ,
\label{eq:t2}
\end{equation}
where the values of the coefficients $C^{R}_{M^\prime B^\prime}$ would depend on the
specific reaction used to excite the resonance and, in this
graphical example, have all been taken to one. The amplitudes have been
calculated for
zero incoming and outgoing relative momentum values. To compare to
actual experiments, one should be using the appropriate excitation coefficients
as well as transition matrix elements going to the on-shell final momentum
corresponding to the value of $\sqrt{s}$. Therefore,
the results in Fig.~\ref{fig:fig2} and similar ones throughout the paper
should be considered mereley illustrative.
The representation is split into the various energy regions where the resonances appear.
In general, a resonance couples dominantly to a given channel and
the value of the maximum of Eq.~(\ref{eq:t2})
is basically proportional to the modulus squared of the product of the
resonance couplings to the dominant and decaying channels, $g_{M^\prime
B^\prime}$ and $g_{M B}$ respectively, and
inversely proportional to the resonance width $\Gamma$.
Note that, instead of adjusting the vertical axis of Fig.~\ref{fig:fig2}b
to the  maximum associated to the narrow resonance at $2805$ MeV,
we have scaled it down to
better visualize the enhancement at 3069 MeV, right below
the $K \Xi_c^\prime$ threshold. This enhancement becomes a resonance if we
increase the cut-off value slightly, with properties that are
similar to the state
found around the same energy by the local models
\cite{Hofmann:2005sw,GarciaRecio:2008dp}.

\subsection{$\Sigma_c$ resonances: (I=1,S=0,C=1) sector}

\begin{table}[ht!]
\begin{center}
\begin{tabular}{|c|c|c||c|c|}
\hline
\multicolumn{5}{|c|}{$(I=1,S=0,C=1)$} \\
\hline
$\Lambda$ [MeV]&\multicolumn{2}{|c||}{ $903$ $(t\ne0)$}&
\multicolumn{2}{|c|}{ $553$ $(t=0)$} \\
\hline $M$ [MeV] & $2551$&$2804$& $2585$&$2804$\\
\hline $\Gamma$ [MeV] &$0.15$ &$5$&$0.005$ &$0.63$\\
\hline
\multicolumn{5}{|c|}{Couplings $|g_i|$} \\
\hline $\pi \Lambda_{c}(2424)$ &$0.06$&$0.27$&$0.002$&$0.04$ \\
\hline $ \pi \Sigma_{c}(2591)$ & $4.00$&$0.16$& $2.15$&$0.04$\\
\hline $DN (2806)$ &$1.25$&$2.10$&$0.38$&$1.70$\\
\hline$K\Xi_{c}(2963) $ &$0.04$&$0.20$ &$0.003$&$0.06$ \\
\hline $\eta\Sigma_{c}(2999) $ &$0.79$&$0.11$&$0.44$&$0.03$\\
\hline $K\Xi'_{c}(3070) $ & $2.30$&$0.14$ & $1.55$&$0.04$\\
\hline $ D_{s}\Sigma(3162)$ &$0.62$&$1.79$&$0.17$&$1.37$\\
\hline $ \eta'\Sigma_{c}(3410)$ &$0.04$& $0.19$&$0.006$& $0.09$\\
\hline $\bar{D}\Xi_{cc}(5307) $ &$0.91$&$0.15$&$0.30$&$0.02$ \\
\hline $\eta_{c}\Sigma_{c}(5432) $ & $0.13$&$0.55$& $0.02$&$0.27$\\
\hline
\end{tabular}
\end{center}
\caption{Masses, widths and couplings of the resonances in the
$(I,S,C)=(1,0,1)$
 sector, for the non-local $(t\ne0)$ and local $(t = 0)$
models.}
\label{tab:tb7}
\end{table}



Using the same cut-off that reproduces the $\Lambda_c(2595)$ in the
$(I,S,C)=(0,0,1)$ sector, we predict two narrow resonances at $2551$
and $2804$ MeV
that appear right below the thresholds of the channels to which they couple
mostly, namely $\pi\Sigma_c$ and $DN$, respectively, as can be seen from
Table~\ref{tab:tb7}. We also show the results obtained with the $t=0$ model
and the same cut-off value of $\Lambda=553$ MeV adjusted to the position of the
$\Lambda_c(2595)$ in the isoscalar channel. As expected, the resonance at 2804 MeV, which couples mostly to $DN$ states, barely changes its position from that of the
non-local model. Indeed, since the $\Lambda_c(2595)$ also couples
mostly to $DN$, the cut-off adjustment of the local model using this resonance as reference has essentially left the $DN$ amplitude intact in this energy region, thereby
generating similar ``DN-type" bound states as in the non-local model in the various isospin sectors.
Note, however, that the width of the resonance at 2804 MeV is
an order of magnitude smaller in the local model.
The lower energy resonance, coupling
mostly to $\pi\Sigma_c$ states, appears at somewhat larger energies in
the local model and, in spite of the gain in phase space, its width is
substantially reduced. This comparison confirms the trend already observed in the
case of the $I=0$ sector. The resonances of the local model appear at similar
energies as those of the non-local approach but their widths are much smaller.
Since this will be a general behavior in all sectors,
we will only show,  in the remaining sections, the results obtained with the
non-local approach developed in this work.

\begin{figure}[ht!]
   \includegraphics[height=7cm,angle=0]{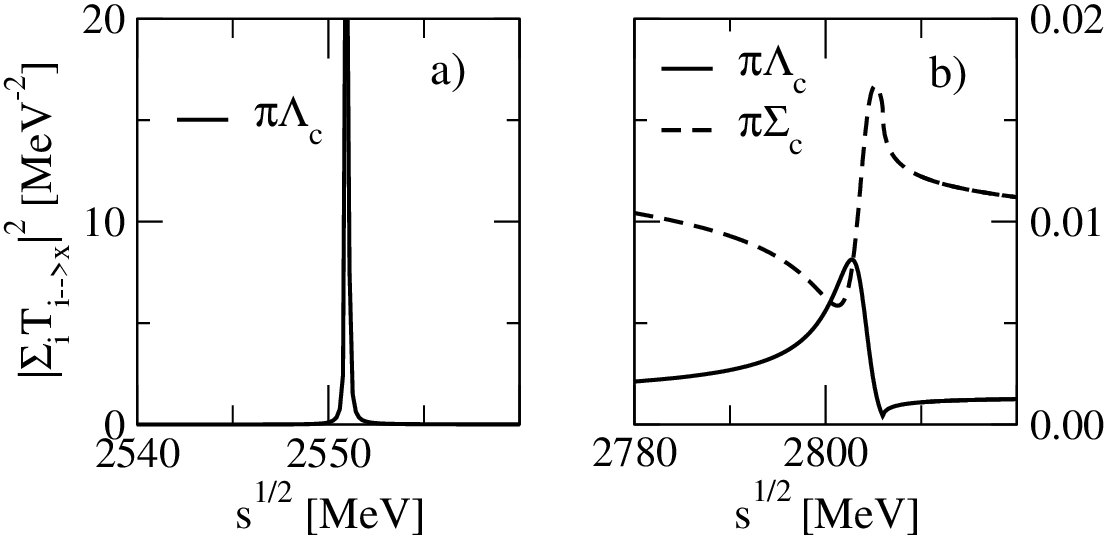}
   \vspace{0.25cm}
   \caption{The same as Fig.~\ref{fig:fig2} but for the
   $(I,S,C)=(1,0,1)$ sector.}
   \label{fig:fig3}
\end{figure}

 The sum of amplitudes squared is represented in
Fig.~\ref{fig:fig3}.
Since the $\Sigma_c(2551)$ resonance couples negligibly to its only allowed
strong
decaying channel, $\pi\Lambda_c$, it is seen in Fig.~\ref{fig:fig3}a as a narrow peak which
makes it easy to miss given the energy resolution of the meson-baryon invariant
masses built up in the present experiments.
The couplings to the different meson-baryon states of the $\Sigma_c$
resonance at 2804 MeV, visualized in Fig.~\ref{fig:fig3}b, allow one to identify it with the state found in Ref.~\cite{Hofmann:2005sw} at a substantially lower
energy, 2680 MeV, using a subtraction method to regularize the loops, as well as that found
in Ref.~\cite{Mizutani:2006vq} around 2750 MeV, using a cut-off method which preserves
isospin symmetry in the regularization scheme. Our result is obviously closer to that of the latter work.

The Belle Collaboration reported recently \cite{Mizuk:2004yu} an isotriplet of excited charmed baryons, $\Sigma_c(2800)$, decaying into $\Lambda_c^+\pi^-$,
$\Lambda_c^+\pi^0$ and $\Lambda_c^+\pi^+$ pairs and having a
width of around $60$ MeV with more than 50\% error. Although this resonance
has been tentatively assumed to decay to $\Lambda_c \pi$ pairs in
d-wave and its spin parity estimated to be $J^P=3/2^-$, actual
angular distributions have not been measured and the fits to $\Lambda_c\pi$ spectra can not rule out s-wave type decays.
Hence, our state at 2805 MeV could be easily identified with the $\Sigma_c(2800)$ resonance, provided three-body decay mechanisms not accounted for in our model, could explain the large width observed experimentally.

\subsection{$\Xi_c$ resonances: (I=1/2,S=-1,C=1) sector}

\begin{figure}[htbp]
   \includegraphics[height=6cm,angle=0]{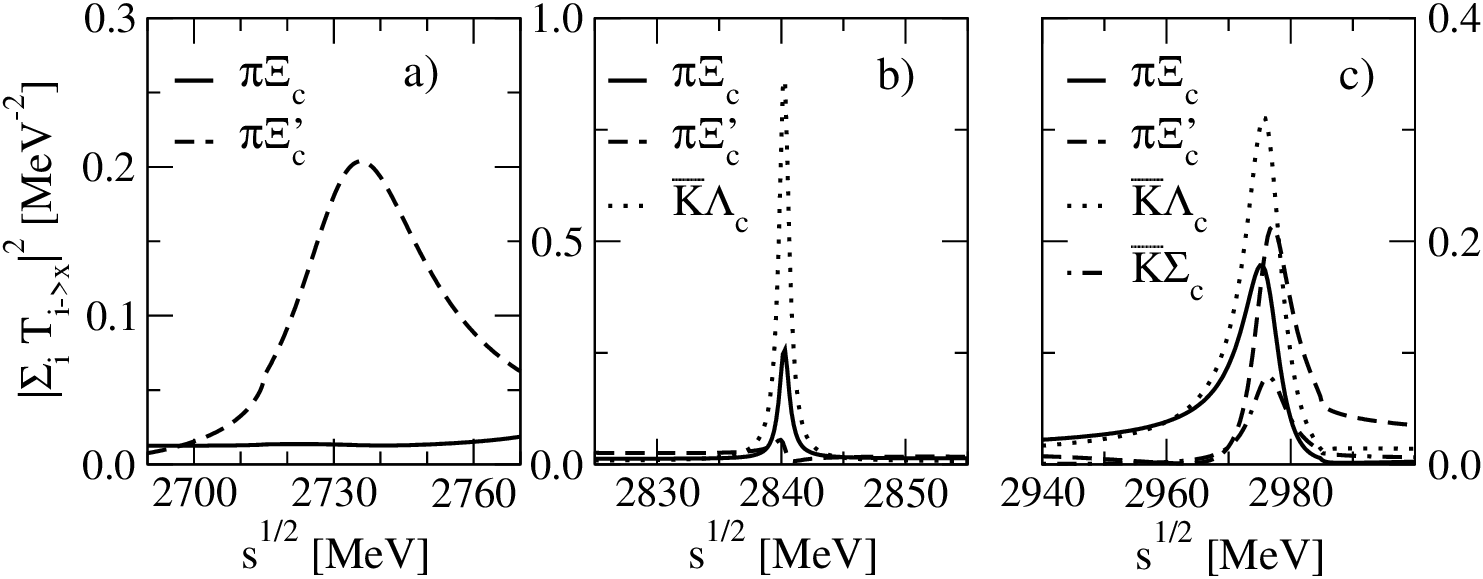}
   \vspace{0.25cm}
   \caption{The same as Fig.~\ref{fig:fig2} but for the
   $(I,S,C)=(1/2,-1,1)$ sector.}
\label{fig:fig25}
\end{figure}

The results in this sector are presented in Table \ref{tab:tb8} and
Fig.~\ref{fig:fig25}.  
We obtain two pure bound
states at $2515$ and $2549$ MeV, respectively, which are placed less than one
pion mass above the $\Xi_c$ member of the $J^P=1/2^+$  ground state
antitriplet, $ (\Lambda_c^+,\Xi_c^+,\Xi_c^0)$, and below the mass of the
$\Xi_c^\prime$ member of the $J^P=1/2^+$ sextet,
$(\Sigma_c^0,\Sigma_c^+,\Sigma_c^{++},\Xi_c^{\prime\,0},\Xi_c^{\prime\,+},\Omega_c^0)$.
This implies that these bound states would decay electromagnetically through
the emission of $\Xi_c \gamma$ pairs and may have been detected at photon
energies of about 50 and 80 MeV in the experiment where the $\Xi_c^\prime$ was
observed \cite{Jessop:1998wt}. Although no apparent signals are reported, we
note that the limited statistical significance of the spectra measured in
\cite{Jessop:1998wt} prevents one from ruling out the existence of these bound
states. Moreover, their production rate would also be somewhat inhibited by the
fact that they are predominantly 5 quark-component states.  Note also that,
lowering gradually the value of the cut-off to somewhat below 700 MeV, the
first state at $2515$ MeV eventually becomes resonance, but quite narrow
due to its weak coupling to the first channel,  whereas the second state at
$2549$ MeV, which is a $\pi\Xi_c$ molecule, rapidly becomes so wide that it
would be difficult to be distinguished from the background.

In addition, our model gives three more resonances above the $\pi\Xi_c$
threshold and below $3$ GeV,  placed at $2733$, $2840$ and $2977$ MeV. The
local model of Ref.~\cite{Hofmann:2005sw}, based on on-shell amplitudes, also obtains three resonances in this
energy region, located in general at  somewhat lower masses and showing a
different order of appearance, as can be inferred from the values of their
couplings to the different meson-baryon components. More specifically, the
lowest resonance in the local model appearing at 2691 MeV and coupling strongly
to $D\Sigma$ should be identified with our middle resonance at 2840 MeV. The
next two resonances appear quite close in the scheme of
Ref.~\cite{Hofmann:2005sw}, at 2793 MeV and 2806 MeV, coupling mostly to
$\bar{K}\Sigma_c$ and $D\Lambda$, respectively, while in our case they are
further apart from each other, at 2733 and 2977 MeV. The crossing in
the ordering of states is another consequence of the
different values of the transition potential amplitudes used in both
coupled-channel schemes.
\begin{table}[htbp]
\begin{center}
\begin{tabular}{|c|c|c|c|c|c|}
\hline
\multicolumn{6}{|c|}{$(I=1/2,S=-1,C=1)$} \\
\hline $M$ [MeV] &$2515$&$2549$ &$2733$ & $2840$ & $2977$  \\
\hline $\Gamma$ [MeV] &$0.$&$0.$ &$34$ & $0.58$ &$4$  \\
\hline
\multicolumn{6}{|c|}{Couplings $|g_i|$} \\
\hline  $\pi \Xi_{c}(2609)$  &$0.65$&$4.47$ &$0.05$ & $0.06$& $0.31$\\
\hline  $\pi \Xi'_{c}(2715)$  &$4.84$&$0.76$ &$1.77$& $0.01$&$0.22$ \\
\hline  $\bar{K}\Lambda_{c}(2779)$ &$0.48$&$3.21$ &$0.19$ &$0.10$ &$0.19$ \\
\hline  $ \bar{K}\Sigma_{c}(2946)$ &$6.90$&$1.01$ &$7.37$ & $0.93$&$0.16$\\
\hline  $D\Lambda (2985)$ &$1.03$& $0.30$& $0.96$& $1.54$& $2.95$\\
\hline  $\eta\Xi_{c} (3018)$ &$0.13$&$1.04$ &$0.13$& $0.18$&$0.10$ \\
\hline  $D\Sigma (3062)$ &$2.91$&$0.89$ & $3.64$& $8.82$ & $1.74$ \\
\hline  $\eta\Xi'_{c}(3124)$ &$4.04$&$0.59$ & $3.47$&$0.46$ & $0.07$ \\
\hline  $K\Omega_{c} (3192)$ &$4.40$&$0.68$ & $1.47$&$0.07$ &$0.21$ \\
\hline  $ D_{s}\Xi(3288)$ &$1.74$&$0.15$ &$0.92$& $4.71$& $2.51$ \\
\hline  $\eta'\Xi_{c}(3428)$ &$0.16$&$0.05$ &$0.15$&$0.53$ &$0.06$ \\
\hline  $\eta'\Xi'_{c}(3534)$ &$0.01$&$0.03$ &$0.09$&$0.11$ &$0.28$ \\
\hline  $\bar{D}_{s}\Xi_{cc}(5408)$ &$0.02$ &$0.01$ &$0.25$ &$0.35$ &$0.71$ \\
\hline  $\bar{D}\Omega_{cc} (5429)$ &$1.10$ & $0.53$ & $1.18$&$0.24$ &$0.03$\\
\hline  $\eta_{c}\Xi_{c}(5450)$ &$1.17$&$1.10$ & $0.55$& $0.04$&$0.03$\\
\hline  $\eta_{c}\Xi'_{c}(5556)$ &$0.04$&$0.08$ &$0.26$ & $0.33$&$0.82$ \\
\hline
\end{tabular}
\end{center}
\caption{Masses, widths and couplings of the resonances in the
$(I,S,C)=(\frac{1}{2},-1,1)$ sector}
\label{tab:tb8}
\end{table}

Recently, several $\Xi_c$ states have been observed by the CLEO
\cite{Csorna:2000hw}, Belle \cite{Chistov:2006zj} and
BaBar \cite{Aubert:2007dt} Collaborations, out of which the three star possible
candidates to be identified with one of our states are at 2790, 2980
and 3080 MeV [see Table~\ref{tab:exp}]. A change of the cut-off value within
a reasonable range could bring any of our two lower mass resonances to agree in
position with the $\Xi_c(2790)$ but the width would turn out to be
twice wider than the observed one in the case of the lower mass state.
The $\Xi_c(2980)$ could be easily associated to either one of the two higher
mass states found here. However, the experimental analysis of Ref.~\cite{Aubert:2007dt}
concludes that the $\Xi_c(2980)^+$ state decays in about 50\% of the cases
into $\Sigma_c^{++} K^-$ pairs, which makes our state at 2840 MeV, showing a
stronger coupling to $\bar{K}\Sigma_c$, the most likely
candidate to be associated to the $\Xi_c(2980)$.

\subsection{$\Omega_c$ resonances: (I=0,S=-2,C=1) sector}

\begin{figure}[ht!]
   \includegraphics[height=7cm,angle=0]{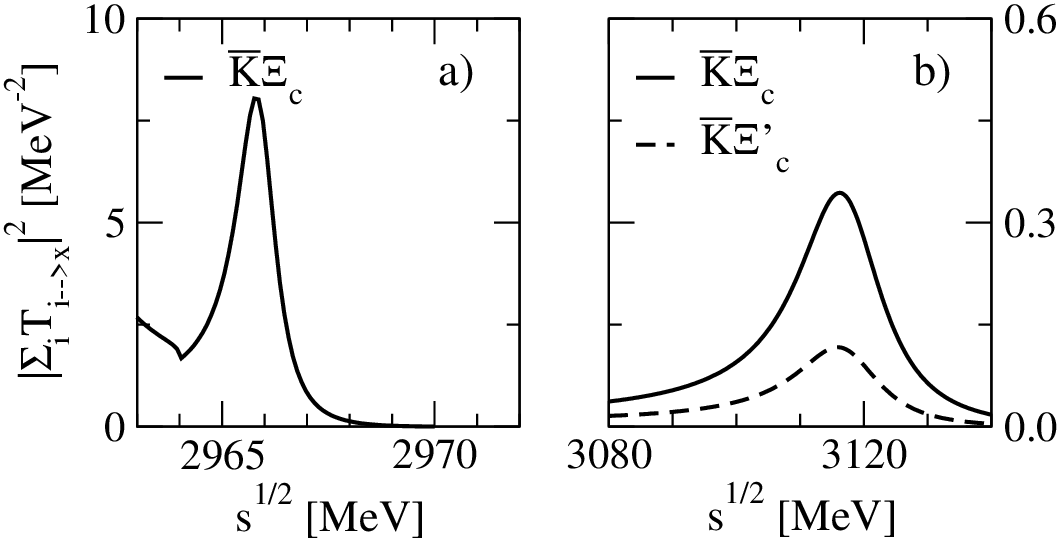}
   \vspace{0.25cm}
   \caption{The same as Fig.~\ref{fig:fig2} but for the
   $(I,S,C)=(0,-2,1)$ sector.}
 \label{fig:fig17}
\end{figure}

In this sector we predict the existence of a bound state at $2959$ MeV, near the lowest
threshold, and two resonances placed at $2966$ and $3117$ MeV  as
can be seen in Table \ref{tab:tb9} and Fig.~\ref{fig:fig17}.
The possible bound state could be detected through the decay into
$\Omega_c \gamma$ states with photons of $E_{\gamma} = 260$ MeV in the
center-of-mass frame. The
resonance placed at $2966$ MeV and seen in Fig.~\ref{fig:fig17}a
is very narrow ($\Gamma=1.1$ MeV) according to the low
coupling of the resonance to the only channel in which it can decay
($\bar{K}\Xi_c$) and the little available phase space.
The resonance at $3117$ MeV with a width of $\Gamma=16$ MeV,
seen in Fig.~\ref{fig:fig17}b, is
a $D\Xi$ molecule that can decay into
$\bar{K}\Xi_c$ and $\bar{K}\Xi'_c$ states.

\begin{table}[htbp]
\begin{center}
\begin{tabular}{|c|c|c|c|}
\hline
\multicolumn{4}{|c|}{$(I=0,S=-2,C=1)$}\\
\hline $M$ [MeV] & $2959$ & $2966$ &$3117$\\
\hline $\Gamma$ [MeV] &$0.$&$1.1$&$16$\\
\hline
\multicolumn{4}{|c|}{Couplings $|g_i|$}\\
\hline ${\bar{K}\Xi_{c}}(2964)$ &$1.36$&$0.43$&$0.51$\\
\hline ${\bar{K}\Xi'_{c}}(3070)$ &$2.04$&$4.49$&$0.27$\\
\hline ${D\Xi}(3189)$ & $2.03$&$1.68$&$5.34$\\
\hline ${\eta\Omega_{c}}(3246)$&$1.67$&$3.69$&$0.24$\\
\hline ${\eta'\Omega_{c}}(3656)$ &$0.10$&$0.07$&$0.35$\\
\hline ${D_{s}\Omega_{cc}}(5528)$ &$0.17$ &$1.17$&$0.19$\\
\hline ${\eta_{c}\Omega_{c}}(5678)$ & $0.28$&$0.21$&$1.03$\\
\hline
\end{tabular}
\end{center}
\caption{Masses, widths and couplings of the resonances in the
$(I,S,C)=(0,-2,1)$ sector}
\label{tab:tb9}
\end{table}

The work of Ref.~\cite{Hofmann:2005sw} also finds three states but placed at
lower energies, 2839, 2928 and 2953 MeV, which follows the trend observed for
other sectors. The pattern of couplings to the various meson-baryon states also
differs a little owing to the different interaction model used. The highest
energy resonance in Ref.~\cite{Hofmann:2005sw}, coupling strongly to
$\bar{K}\Xi_c^\prime$ and $\eta\Omega_c$, would correspond to our middle one,
while the lowest one in  Ref.~\cite{Hofmann:2005sw}, coupling strongly to
$D\Xi$, would be the equivalent to our resonance at higher energy.

\subsection{Resonances of five quarks}

We have also analyzed the sectors corresponding to resonances that cannot be
realized with only three quarks and, therefore, their existence would be
signaling the presence of pure five quark states.
Note that the possible pentaquark-type systems predicted by the present
model would be color singlet states built up from combinations of
color singlet $q{\bar q}$ with color singlet $qqq$ components. States with a different composition, such as a combination of color octet $q{\bar q}$ with color octet $qqq$ clusters,
can not be generated by our meson-baryon scattering approach.

\begin{figure}[htbp]
   \includegraphics[height=7cm,angle=0]{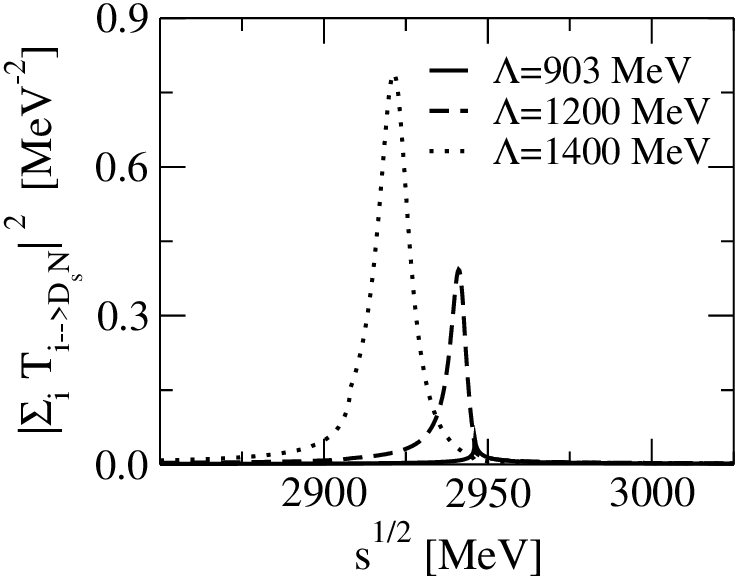}
   \vspace{0.25cm}
   \caption{The same as Fig.~\ref{fig:fig2} but for the
   $(I,S,C)=(1/2,1,1)$ sector and for three different cut-off values:
    $903$, $1200$ and $1400$ MeV.}
   \label{fig:fig15}

\end{figure}

\begin{table}[htbp]
\begin{center}
\begin{tabular}{|c|c|c|c|}
\hline
\multicolumn{4}{|c|}{$(I=1/2,S=1,C=1)$}\\
\hline $\Lambda$ [MeV]&$903$&$1200$&$1400$\\
\hline $M$ [MeV] & $2946$&$2941$&$2924$\\
\hline $\Gamma$ [MeV] &$0.93$&$5$&$12$\\
\hline
\multicolumn{4}{|c|}{Couplings $|g_i|$}\\
\hline ${K\Lambda_c}(2779)$ & $0.002$ & $0.04$ & $0.10$ \\
\hline ${D_s N}(2908)$ & $0.03$ & $0.84$ & $1.68$ \\
\hline ${K\Sigma_c}(2946)$ & $0.07$ & $1.79$ & $3.59$ \\
\hline
\end{tabular}
\end{center}
\caption{Masses, widths and couplings of the resonances in the
$(I,S,C)=(1/2,1,1)$ sector for different cut-off values: $903$, $1200$ and $1400$ MeV.}
\label{tab:tb11}
\end{table}

Out of the three possible
sectors, namely $(I,S,C)=(2,0,1)$, $(\frac{1}{2},1,1)$ and $(\frac{3}{2},1,1)$,
we only find hints of a possible resonance in the case $I=1/2,S=1,C=1$, where we
see a cusp-like structure placed at the
threshold of the $K\Sigma_{c}$ channel to which the state couples more
strongly. This behavior is shown by the solid line in Fig.~\ref{fig:fig15}
and by the first column of coupling constants displayed in Table~\ref{tab:tb11},
which have been obtained using our nominal cut-off value of 903 MeV.
According to the mechanism discussed in Ref.~\cite{Toki:2007ab}, the
coupling constants should vanish if the resonance was placed right at the
$K\Sigma_{c}$ threshold, which explains the smallness of their values.
In order to
see whether the cusp structure  would eventually become a clear resonance with
a slight change of parameters,  we also display in Fig.~\ref{fig:fig15} and in
Table~\ref{tab:tb11} our results with two other values  of the cut-off, $1200$
and $1400$ MeV. One can clearly see that the cusp structure becomes a more
bound and wider resonance as the cut-off value increases, while the coupling
constants become larger.

\subsection{Dependence on model parameters}

We finalize this work by exploring the dependence of our results on the shape
and size of the form factor employed, which are ingredients of the model that
are not constrained by symmetry arguments.

First, replacing the dipole-type form
factor by a gaussian form:
$$F(\mid \vec{k}\mid)={\rm e}^{-\frac{\vec{k}^2}{2\Lambda_g^2}} \ ,$$
we are able to adjust the position of the $\Lambda_c(2595)$ with a gaussian
cut-off value of
$\Lambda_g=543$ MeV. The corresponding width is exactly the same as that found for the dipole-type form factor. When exploring the other sectors, the gaussian form factor gives rise to the same resonances, some of them slightly displaced by at most 50 MeV from the position found with the dipole-type form factor, but having essentially the same width.

Retaining the dipolar form factor, we next explore the effects of
varying the value of the cut-off $\Lambda$, a parameter that is not constrained by symmetry arguments, between 600 and 1200 MeV, that is, up to 300 MeV below
and above the nominal value of 903 MeV used in this work. This variation
produces changes in the positions and widths of the resonances within certain
ranges, the general trend of which are summarized in the following points:
 \begin{enumerate}
 \item A resonance that lies far below --by 50 to 200 MeV-- the meson-baryon
 threshold to which it couples more strongly may change its position by an
 amount comparable with the variation of the cut-off value. The larger the
 cut-off the more bound the resonance becomes.
 \item The width of the resonance only changes appreciably for cut-off values
 that move the resonance above the threshold of a meson-baryon channel to which
 the resonance couples significantly.
 \item Weakly bound resonances change their positions more moderately, at most
by 10 MeV for changes of cut-off values within 100 MeV. In this case, the width
tends to decrease as the resonance becomes less bound because of the
distortions induced by moving closer to the threshold,
a phenomenon also known as Flatt\'e effect \cite{flatte}.
\end{enumerate}

Having explored the systematics to the cut-off changes, we finally summarize in
Table~\ref{tab:summary} the states which, taking an appropriate
cut-off value within the range explored,  could be identified with a well
established resonance of $J^P=1/2^-$ or unknown spin-parity.
\begin{table}[htbp]
     \setlength{\tabcolsep}{0.2cm}
\begin{center}
\begin{tabular}{|c|c|ccc|ccc|}
\hline
$(I,S)$ & $\Lambda$ [MeV] & \multicolumn{3}{|c|}{Theory} &
\multicolumn{3}{|c|}{Experiment}\\
 & & Mass [MeV] & main channel & Width [MeV] & Mass [MeV] & Width [MeV] & Status \\
\hline
\hline
$(0,0)$ & 903 & 2595 & $DN(2806)$ & 0.5 & $2595.4 \pm 0.6$ & $3.6^{+2.0}_{1.3}$ & *** \\
$\Lambda_c$ &  &  &  &  & &  & \\
\hline
$(1,0)$ & 1100 & 2792 & $DN(2806)$ & 16 &  $2801^{+ 4}_{-6}$ & $75^{+22}_{-17} ~~(\Sigma_c^{++})$ & \\
$\Sigma_c$ & & & & & $2792^{+ 14}_{- 5}$ & $62^{+60}_{-40}~~(\Sigma_c^+)$ & *** \\
 & & & & &  $2802^{+4}_{- 7}$ & $61^{+28}_{-18}~~(\Sigma_c^0)$ & \\
 \hline
$(\frac{1}{2},-1)$ & 814 & 2790 & ${\bar K}\Sigma_c(2946)$ & 55 & $2789.1 \pm 3.2$ & $<15 ~~(\Xi_c^+)$ & *** \\
$\Xi_c$ & 980 & 2790 & $D\Sigma(3062)$ & 0.5 & $2791.8 \pm 3.3$ &
$<12 ~~(\Xi_c^0)$ & \\
&  & \multicolumn{3}{|c|}{} &
\multicolumn{3}{|c|}{}\\
 & 655 & 2970 & $D\Sigma(3062)$ & 1.2 & $2971.4 \pm 3.3$ & $26\pm 7~~(\Xi_c^+)$ & *** \\
 & 960 & 2970 & $D\Lambda(2980)$ & 5.1 & $2968.0 \pm 2.6$ & $20\pm 7~~(\Xi_c^0)$ & \\
\hline
\end{tabular}
\end{center}
\caption{Masses, widths and main coupled channel of states that can be
identified with well established resonances in various sectors.}
\label{tab:summary}
\end{table}

\section{Conclusions}
\label{sec:Conclusions}

We have studied charmed baryon resonances obtained from a coupled
channels unitary approach using a t-channel vector-exchange driving
force.

To the best of our knowledge, all
previous models of dynamically generated baryon resonances in the charm sector
rely on a local zero-range interaction, which is
obtained by neglecting the four-momentum transfer $t$
in front of the mass of
the exchanged vector meson squared, $m_V^2$.  However,
we have illustrated that the value $t/m_V^2$ is not at all negligible in the
heavy sector, especially for charm-exchange
amplitudes which produce a large value of the four-momentum transfer due
to the large difference bewteen the masses of the mesons involved in the
transition.

We have analyzed in detail the effects of going beyond the $t=0$
approximation and, taking the
 $I=0$, $C=1$, $J=1/2$ sector of the well established
$J=1/2^-$ $\Lambda_c(2595)$ resonance as reference, we find that the
experimental data is better reproduced by the non-local model which also
requires a more reasonable cut-off regularization value of 903 MeV.

Compared to the local models based on on-shell amplitudes, our approach obtains basically the same amount of
resonances in all sectors but appearing, in general,
at somewhat larger energies because the diagonal amplitudes,
largely responsible for generating the bound states, are smaller in magnitude.
An essential finding of this work is that our non-local approach produces much
wider resonances because of the larger value
of the non-diagonal amplitudes when $t\neq 0$.

Varying the cut-off parameter within a reasonable range, we are able to locate
some of our states at the energy position of a measured resonance in the same
sector. In particular, we suggest
the identification of the $\Lambda_c(2595)$, $\Sigma_c(2800)$, $\Xi_c(2790)$ and
$\Xi_c(2980)$ as dynamically generated resonances having $J^P=1/2^-$.
In general, the widths of
the states produced by our model are smaller than the
experimentally observed ones, since we do not account for three-particle
decay channels.

We find a possible resonance in the sector with quantum numbers
$(I,S,C)=(\frac{1}{2},1,1)$
that can only be realized by the consideration of a minimum of five quarks.
The cusp-like structure observed at
the threshold of the $K\Sigma_{c}$ channel for a cut-off value of 903 MeV,
becomes a more
bound and wider clear resonance as the cut-off value increases.

This is the first exploratory study of the effects tied to the non-locality of
the meson-baryon interaction in the charm sector. We have considered
meson-baryon coupled states built up from the $J^P=0^-$
mesons and the ground state $J^P=1/2^+$ baryons. However, Heavy Quark
Symmetry demands
that the states containing heavy vector mesons are treated on equal footing due
to the similarity of the masses with the heavy pseudoscalar ones. This will be
addressed in a future work.


\section*{Acknowledgements}
We are very grateful to M. F. M. Lutz for helpful discussions and comments.
We also thank E. Graug\'es for clarifications on the details of the experiments,
 and B. Juli\`a--D\'{i}az, V.K. Magas and A. Parre\~no for interesting
 suggestions and their help in various stages of the calculation.
This work is partly supported by the EU contract No. MRTN-CT-2006-035482
(FLAVIAnet), by the contract FIS2008-01661 from MIC
(Spain), by the Ge\-ne\-ra\-li\-tat de Catalunya contract 2009SGR-1289,
and by FEDER/FCT (Portugal) under the
project CERN/FP/83505/2008. We
acknowledge the support of the European Community-Research Infrastructure
Integrating Activity ``Study of Strongly Interacting Matter'' (HadronPhysics2,
Grant Agreement n. 227431) under the Seventh Framework Programme of EU.


\end{document}